\documentclass[twocolumn,showpacs,preprintnumbers,amsmath,amssymb]{revtex4}

\usepackage{graphicx}
\usepackage{dcolumn}
\usepackage{bm}

\begin{document}


\title{THz radiation by the frequency down-shift of Nd:YAG lasers}

\author{S. Son}
\affiliation{18 Caleb Lane, Princeton, NJ 08540}
\author{Sung Joon Moon}
\affiliation{28 Benjamin Rush Lane, Princeton, NJ 08540}
\author{J.~Y. Park}
\affiliation{Los Alamos National Laboratory}
\date{\today}

\begin{abstract}
The interaction between an intense laser and a relativistic dense electron beam
propagating in the {\em same} direction could down-shift
the laser frequency.
This process, which can be used to generate a coherent THz radiation,
is theoretically analyzed.
With a set of practically relevant parameters,
it is suggested  that the radiation energy could reach the order of $1~\mathrm{mJ}$ per shot
in the duration of 100 pico-second, or the temporal radiation power of 10 MW.
\end{abstract}


\maketitle

A THz light has beneficial characteristics for a range of
applications~\cite{diagnostic, siegel, siegel2,  siegel3},
as it passes through various non-conducting materials while it remains
non-ionizing and non-invasive because of its relatively low photon energy.
Most of the recent research efforts have been focused on the light source of
1 to 10 THz frequency range~\cite{siegel, siegel2, booske}. 
However, the currently available technologies~\cite{gyrotron3, tgyro,qlaser, qlaser3,colson, Free2} 
still have limitations in generating an \textit{intense} and \textit{short} THz light source~\cite{booske}.
On the other hand, great advances have been made in developing ultra-short and high power
energy sources,
in the context of the inertial confinement fusion~\cite{Fisch, sonbackward, sonlandau,tabak,sonprl,sonpla}.
The density of the relativistic electron beam can be achieved up to
$10^{20} \mathrm{cm^{-3}}$~\cite{monoelectron, ebeam},
and a $1 \ \mu\mathrm{m}$-light source of the intensity
exceeding $10^{18} \ \mathrm{W} / \mathrm{cm}^2$ with the duration of pico-seconds 
are currently available.
We note that an intense, coherent THz light source could be generated
if the frequency of the above-mentioned lasers gets properly shifted
to the THz regime.
There have been successful attempts to generate the THz radiation using the laser driven electron beams~\cite{Carr, Thz, Thz2}.

In this paper, one frequency down-shifting scheme is theoretically investigated,
which is inspired by the principle of the free electron laser (FEL) utilizing an intense
electron beam and an appropriate laser.
In the conventional laser-based FEL, a well-collimated relativistic
electron beam propagates in the {\em opposite} direction to the intense laser. 
The amplified wave of the frequency $\omega = 2 \gamma_0^2 \omega_0$ 
gets radiated in the same direction with the electron beam,
where $\omega_0$ is the original laser frequency and 
$\gamma_0$ is the relativistic factor of the beam,
a function of the beam speed and the speed of light $c$.
On the contrary,
if both the relativistic electron beam and the laser propagate in the {\em same} direction, 
the FEL mechanism would result in the wave with the {\em down-shifted} frequency of
$\omega = \omega_0/ 4 \gamma_0^2$, propagating in the opposite direction.
Our analysis based on the practically relevant parameters
reveals that the THz wave with the energy more than 1 mJ could be radiated per shot.
In the rest of the paper, we study the interaction between the co-traveling
relativistic electron beam and an intense laser, estimate the amplification
efficiency, and then discuss the attractive features of this scheme. 

Let us consider an intense laser and a relativistic electron beam, both traveling in
the positive $z$-direction.
We assume the laser is linearly polarized, given by
$\mathbf{E} = E_0 \cos( k_0z -\omega_0t) \hat{x}$ 
and $\mathbf{B} =  E_0 \sin( k_0z - \omega_0t) \hat{y}$.
In this paper, we assume that  $\omega_0 / k_0 = c $ so that 
$|\mathbf{E}| = |\mathbf{B}|$,  which is valid 
when the electron beam density is much lower than  the critical density of
the laser. This is the case for our regime of interest. 
For simplicity, we ignore the electron motion in $y$-direction. 
The motion of a relativistic electron is given as 
\begin{equation}
m_e \frac{d \gamma(v) \mathbf{v}}{dt}= -e\left[ \mathbf{E} + \frac{\mathbf{v}}{c} \times  \mathbf{B} \right] \mathrm{,}
\end{equation}
where $\gamma(v)^{-1}  = \sqrt{1-(v/c)^2} $ and $v \equiv \mathbf{|v|}$.  
The first order perturbed motion of an electron with the initial velocity $\mathbf{v} = v_0 \hat{z}$ 
is given as 
\begin{equation}
v^{(1)}_{x} = -  \left(\frac{eE_0}{m \omega_0 \gamma_0 }\right) \sin(k_0 z - \omega_0 t), \label{eq:10} 
\end{equation}
where $\gamma_0^{-1}  = \sqrt{1-(v_0/c)^2} $  and the approximations of
$ | \mathbf{E}+ \mathbf{v}/c \times \mathbf{B} | 
\cong  (1- v_0/c) |\mathbf{E}|$ and 
$ k v_0 t - \omega_0 t \cong k (v_0 - c) t$ are used. 
In driving Eq.~(\ref{eq:10}),  it is assumed that  $v^{(1)}_{x}/c < 1$. 
For an intense laser, there is a time dependent velocity (and therefore $\gamma$) along the direction of the propagation (the z-direction); the electron experiences the eight figure motion in the x and z direction. 
For $v^{(1)}_{x}/c <  1 $, this motion will be time-averaged out so that 
 it can be ignored. However,  
the general treatment for the case when $v^{(1)}_{x}/c\cong 1$  is beyond the scope of this paper.

Now, consider a counter-propagating electromagnetic wave given by 
$\mathbf{E} = \delta E_T\sin(k_T z + \omega_T t)\hat{x} $ and $\mathbf{B} = \delta E_T\sin(k_T z + \omega_Tt)\hat{y} $.
We also assume $\omega_T / k_T = c$. 
Since $\omega_t \ll \omega_0 $, the dispersion relationship of the THz wave could be significantly different  from the one in the vacuum for a high electron density beam. 
This diviation of the vacuum dispersion relationship would change the prediction of the resonant frequency. 
However, the qualitative conclusion of our analysis does not depend 
on the detailed dispersion relationship of the THz radiation,
 which will be discussed further in Eq.~(\ref{eq:res2}). 
The second order electron motion is described by
\begin{eqnarray}
\frac{dv^{(2)}_{z}}{dt} =&+& \left(\frac{e \delta E_T}{ m}\right)  \left(\frac{eE_0}{m\omega_0 c}\right) \frac{1}{\gamma_0^4(v_0/c)^2 + \gamma_0^2  } \times \nonumber \\  \nonumber \\
 & &  \sin(k_0 z - \omega_0 t) \sin(k_T z + \omega_T t)
 \label{eq:3} \mathrm{,} 
\end{eqnarray}
where it is assumed that $ v^{(1)}_x \ll v_0$ so that 
 $d(\gamma v_z)/dt \cong (\gamma^3/c^2) v_z^2 (dv_z/dt) + \gamma (dv_z/dt)$.
Considering only the resonance term, Eq.~(\ref{eq:3}) can be simplified to the form of the
FEL pendulum equation
\begin{eqnarray} 
\frac{dv^{(2)}_{z}}{dt} \cong  & &  \kappa(E_0, v_0, k_0) \left(\frac{ e \delta E_T}{ m_e}\right) \nonumber \times \nonumber \\
&&\frac{ \cos\left(\phi_0 + (k_0+k_T)z - (\omega_0 - \omega_T ) t\right)}{2}  \label{eq:pendulum} \mathrm{,}
\end{eqnarray}  
where $ \phi_0$ is the initial phase,
 and  $\kappa(E_0, k_0, v_0) $ is given as 
\begin{equation}
\kappa(E_0, k_0, v_0) =\left( \frac{eE_0}{m \omega_0 c}\right) \frac{1}{\gamma_0^4 (v_0/c)^2 +\gamma_0^2} \label{eq:kappa} \mathrm{.}
\end{equation}
The resonance condition of the pendulum equation, Eq.~(\ref{eq:pendulum}),
is given to be 
\begin{equation} 
k_T  =k_0(c - v_0) / (c + v_0)  \cong k_0 /4 \gamma_0^2  \label{eq:res} \mathrm{.}
\end{equation} 
If the THz dispersion relationship is given as $
\omega_T(k_T)^2 = 4 \pi n_b e^2 / m_e\gamma_0 + c^2 k_T^2 $ deviating from the vacuum, where $n_b$ is the electron beam density,  the resonant condition is 

\begin{equation} 
\left[\frac{v_0}{c} + \sqrt{ \frac{\omega_{\mathrm{bpe}}^2 }{ (ck_T)^2 \gamma_0 } + 1 } \right] k_T = \left(1-\frac{v_0}{c} \right)k_0 \cong \frac{k_0 }{2\gamma_0^2} \mathrm{,} 
\label{eq:res2}
\end{equation}  
where $\omega_{\mathrm{bpe}}^2 = (4 \pi n_b e^2/ m)$ and we need solve for  for $k_T$.  
The exact resonant condition will be different from Eq.~(\ref{eq:res}), but 
  the down-shift will be the same order as in the case when $c k_T = \omega_T$. 
As mentioned,  the qualitative conclusion will not change in any significant way. 

Equating the energy dissipation rate of the  electrons to the wave growth rate,
\begin{equation}
\gamma^i \frac{\delta E_T^2}{4\pi} = n_e (m_e/2)\langle  \gamma_0^3 \frac{d \delta (v^{(2)}_{z})^2}{ dt}\rangle \mathrm{,}
\end{equation}
leads to the Landau damping instability rate (i.e., the FEL growth rate)~\cite{stix, songamma}: 
\begin{equation}
  \gamma^i =  \frac{\pi}{2} \gamma_0^3  \frac{ \kappa(E_0, k_0, v_0)^2 \omega_{\mathrm{bpe}}^2}{ q^2} \left(\frac{\partial f_e}{\partial v}\right)_{\omega / q} \label{eq:growth} \omega \mathrm{,}
\end{equation}
where $\langle \rangle$ is the ensemble average over the electron distribution,
$q = k_0 + k_T \cong k_0 $,  and $\omega = \omega_0 - \omega_T \cong \omega_0$,  $\omega^2_{\mathrm{bpe}} = 4 \pi n_b e^2 / m_e$ 
is the Langmuir frequency of the beam density $n_b$,
and $f_e $ is the electron distribution function 
with the normalization  condition
$\int f_e dv_z = 1$. 
The Landau damping theory suits better than the conventional linear FEL analysis~\cite{colson} 
when the electron beam has high energy spread, as it is the case here~\cite{monoelectron, ebeam, wake, wake2}. 

Eq.~(\ref{eq:growth}) describes the FEL amplification 
of the down-shifted electromagnetic wave.  
The physical mechanism can be understood better by considering the electron
motion in the reference frame where the electron beam is nearly stationary.
In this reference frame, 
the laser is seen propagating in the positive $z$-direction,
with the down-shifted frequency of $\omega_b = \omega_0/2 \gamma_0$. 
If the laser duration in the laboratory frame is $\tau$,
the duration in this frame is $2\gamma_0 \tau$. 
The electron would experience the quiver motion by the laser,
which results in the electromagnetic wave propagating in the opposite direction
(negative $z$-direction) unstable. 
This unstable mode would be 
amplified and emitted in the negative $z$-direction
with the frequency of $\omega = \omega_b / 2 \gamma_0$ 
and the duration  of $2 \gamma_0 \tau $.
Seen in the laboratory frame, it propagates in the negative $z$-direction
with the duration of $4 \gamma_0 ^2\tau$ 
and the further shifted-down frequency 
of $\omega = \omega_b /2 \gamma_0 = \omega_0/4\gamma_0^2$.
For instance,  if the wavelength of the original laser is 1 $\mu \mathrm{m}$ 
and the duration is 1 pico-second,   
the wavelength and the duration of 
the amplified wave would be
100 $\mu \mathrm{m}$ and 100 pico-second for $\gamma_0 = 5$. 

Let us estimate the efficiency of our FEL scheme. 
As an example,
consider 1-$\mu \mathrm{m}$ Nd:YAG laser with the intensity $I$ and the duration $\tau$.  
Denoting $I_{18} = I/ (10^{18} \mathrm{W} / \mathrm{cm}^2$) and 
the energy spread of the electron beam as $\delta E / E = \zeta $,
a crude approximation leads to  
$(\partial f / \partial v)_{\omega/q}  \cong (1/ c^2) (\gamma_0^4/\zeta^2)$. 
As  $\kappa$ in Eq.~(\ref{eq:kappa}) can be estimated to be
$ \kappa^2 = (1/10 \gamma_0^8) I_{18}$,  
Eq.~(\ref{eq:growth}) can be re-casted into 
\begin{eqnarray}
 \gamma^i \tau  &=& 0.17 \times \frac{I_{18}}{\gamma_0}\frac{\omega_{\mathrm{bpe}}^2}{\omega_0^2}
 \frac{1}{\zeta^2} \omega_0 \tau \nonumber \\ \nonumber \\
&=& 0.456   \frac{I_{18} \tau_{-12} n_{18}}{\gamma_0 \zeta^2} \mathrm{,} \label{eq:sim}
\end{eqnarray}
where $\tau_{-12} = \tau / (10^{-12} \sec)$, and $n_{18} = n_b / (10^{18} \mathrm{cm^{-3}})$.
For a simple estimation of the plausible physical parameter, 
let us assume that the electron
beam duration is the same as that of the laser and 
that  $\tau_{-12} = 1$, $\zeta = 0.01$ and $\gamma_0=5.0$.
Then, the condition $\gamma^i \tau > 1$ is given as $I_{18} n_{18} > 0.003$, 
and  the radiation frequency is  3 THz. 

Consider two examples to highlight two competing aspects of this scheme.  
In the first case, assume
$I_{18} = 0.001$, $n_{18} = 10$, $\gamma_0=5.0$ and $\zeta = 0.01$,
which satisfies $\gamma^i \tau > 1$. 
Assuming the spot size of both the electron beam and the laser are roughly 0.01 cm, 
the total energy of the electron and laser beam is 0.1  J and 0.1 J, respectively.
The possible total radiation of the THz radiation per shot would be roughly  0.001 J, 
assuming 1\% of the beam energy ($\zeta = 0.01$) is converted into the THz radiation.   
The radiation time is 100 pico-second and the frequency is 3 THz.  
In the second example, assume $I_{18} = 1 $,
$n_{18} = 0.01$, $\gamma_0 = 5.0$ and  $\zeta = 0.01$, which also satisfies the condition
$\gamma^i \tau > 1$. 
For the same spot size and the duration as in the first case, 
the total energy of the electron beam is 0.1 J and the total energy of the laser beam is 100 J;
1 $\mu \mathrm{J}$ can be radiated in the THz frequency,
assuming the same conversion rate of 1\%.
The radiation time is 100 pico-second and the frequency is 3 THz.

Two examples given above illustrates the following.
For a denser electron beam, the total radiated energy is higher and 
the requirement for the laser intensity is less severe.  
In general, a denser electron beam is preferred. 
However, a denser beam
would expand very fast via the space-charge effect.
In order to avoid this, 
the beam needs to be designed to propagate through a plasma of  
comparable electron density to the beam. 
However, if the background plasma density is too high, exceeding the order of
$10^{17} \mathrm{cm^{-3}} $, the resulting THz radiation would not propagate any more.
It gets reflected toward the inside
the beam and would not propagate through the beam because the THz radiation cannot (can)
penetrate the beam propagating in the same (opposite) direction.   
This trapping mechanism reduces the total energy radiated in the THz frequency.
The optimal parameter regime would be such that 
the trapping of the THz radiation by the background plasma is not severe,
while the electron beam is still dense enough for
 mitigated intensity requirement of the laser. 
Our rough estimate suggests  that 
the energy of the laser and the electron beam is a few J 
and the electron beam density (comparable to that of the background plasma)
is $n_e \cong 10^{18\sim19} \mathrm{cm^{-3}}$.
A more detailed calculation of the optimal regime is the topic of our future research.

To summarize, a scheme for the coherent THz radiation,
based on the frequency down-shift of the FEL where both the 
electron beam and the laser propagate in the same direction,
is theoretically investigated and its practicality is examined.
In this scheme, the laser frequency would be down-shifted by a factor of $4 \gamma^2$,
and the total radiated power is between 1 $\mu \mathrm{J}$ and 1 mJ, depending on the laser intensity
and the electron beam density.
The temporal power may reach as high as  100 MW. 

The scheme proposed has many disadvantages. Firstly, the required laser intensity is high.  Secondly, the required electron beam density is also  high. The space charge effect could be detrimental to the maintenance of the beam quality for sufficiently long time. 
Thirdly, the total radiated power is not as large as in the case of the gyrotron.  Lastly, there are many physical effects that we do not consider, which could prevent the scheme from being practical. 
The time dependent electron motion in the z-direction and the modified dispersion relationship of the radiated THz wave are among those physics. However,  as the laser technologies develop in a very fast phase,  the current scheme could get more practical since the major obstacles of the conventional FEL, such as the expensive magnets, can be avoided. 

\bibliography{tera2}

\begin{thebibliography}{26}
\expandafter\ifx\csname natexlab\endcsname\relax\def\natexlab#1{#1}\fi
\expandafter\ifx\csname bibnamefont\endcsname\relax
  \def\bibnamefont#1{#1}\fi
\expandafter\ifx\csname bibfnamefont\endcsname\relax
  \def\bibfnamefont#1{#1}\fi
\expandafter\ifx\csname citenamefont\endcsname\relax
  \def\citenamefont#1{#1}\fi
\expandafter\ifx\csname url\endcsname\relax
  \def\url#1{\texttt{#1}}\fi
\expandafter\ifx\csname urlprefix\endcsname\relax\def\urlprefix{URL }\fi
\providecommand{\bibinfo}[2]{#2}
\providecommand{\eprint}[2][]{\url{#2}}

\bibitem[{\citenamefont{Nagel et~al.}(2001)\citenamefont{Nagel, Bolivar,
  Brucherseifer, Kurz, Bosserhoff, and Buttner}}]{diagnostic}
\bibinfo{author}{\bibfnamefont{M.}~\bibnamefont{Nagel}},
  \bibinfo{author}{\bibfnamefont{P.~H.} \bibnamefont{Bolivar}},
  \bibinfo{author}{\bibfnamefont{M.}~\bibnamefont{Brucherseifer}},
  \bibinfo{author}{\bibfnamefont{H.}~\bibnamefont{Kurz}},
  \bibinfo{author}{\bibfnamefont{A.}~\bibnamefont{Bosserhoff}},
  \bibnamefont{and} \bibinfo{author}{\bibfnamefont{R.}~\bibnamefont{Buttner}},
  \bibinfo{journal}{Appl.~Phys.~Lett.} \textbf{\bibinfo{volume}{80}},
  \bibinfo{pages}{154} (\bibinfo{year}{2001}).

\bibitem[{\citenamefont{Siegel}(2002)}]{siegel}
\bibinfo{author}{\bibfnamefont{P.~H.} \bibnamefont{Siegel}},
  \bibinfo{journal}{Microwave Theory and Techniques, IEEE Transaction on}
  \textbf{\bibinfo{volume}{50}}, \bibinfo{pages}{910} (\bibinfo{year}{2002}).

\bibitem[{\citenamefont{Siegel}(2004)}]{siegel2}
\bibinfo{author}{\bibfnamefont{P.~H.} \bibnamefont{Siegel}},
  \bibinfo{journal}{Microwave Theory and Techniques, IEEE Transaction on}
  \textbf{\bibinfo{volume}{52}}, \bibinfo{pages}{2438} (\bibinfo{year}{2004}).

\bibitem[{\citenamefont{Siegel}(2007)}]{siegel3}
\bibinfo{author}{\bibfnamefont{P.~H.} \bibnamefont{Siegel}},
  \bibinfo{journal}{Antennas and Propagation, IEEE Transactions on}
  \textbf{\bibinfo{volume}{55}}, \bibinfo{pages}{2957} (\bibinfo{year}{2007}).

\bibitem[{\citenamefont{Booske}(2008)}]{booske}
\bibinfo{author}{\bibfnamefont{J.~H.} \bibnamefont{Booske}},
  \bibinfo{journal}{Physics of Plasmas} \textbf{\bibinfo{volume}{15}},
  \bibinfo{pages}{055502} (\bibinfo{year}{2008}).

\bibitem[{\citenamefont{Bratman et~al.}(2009)\citenamefont{Bratman, Kalynov,
  and Manuilov}}]{gyrotron3}
\bibinfo{author}{\bibfnamefont{V.~L.} \bibnamefont{Bratman}},
  \bibinfo{author}{\bibfnamefont{Y.~L.} \bibnamefont{Kalynov}},
  \bibnamefont{and} \bibinfo{author}{\bibfnamefont{V.~N.}
  \bibnamefont{Manuilov}}, \bibinfo{journal}{Phys.~Rev.~Lett.}
  \textbf{\bibinfo{volume}{102}}, \bibinfo{pages}{245101}
  (\bibinfo{year}{2009}).

\bibitem[{\citenamefont{Glyavin et~al.}(2008)\citenamefont{Glyavin, Luchinin,
  and Golubiatnikov}}]{tgyro}
\bibinfo{author}{\bibfnamefont{M.~Y.} \bibnamefont{Glyavin}},
  \bibinfo{author}{\bibfnamefont{A.~G.} \bibnamefont{Luchinin}},
  \bibnamefont{and} \bibinfo{author}{\bibfnamefont{G.~Y.}
  \bibnamefont{Golubiatnikov}}, \bibinfo{journal}{Phys.~Rev.~Lett.}
  \textbf{\bibinfo{volume}{100}}, \bibinfo{pages}{105101}
  (\bibinfo{year}{2008}).

\bibitem[{\citenamefont{Faist et~al.}(1994)\citenamefont{Faist, Capasso, Sivco,
  Sirtori, Hutchinson, and Cho}}]{qlaser}
\bibinfo{author}{\bibfnamefont{J.}~\bibnamefont{Faist}},
  \bibinfo{author}{\bibfnamefont{F.}~\bibnamefont{Capasso}},
  \bibinfo{author}{\bibfnamefont{D.~L.} \bibnamefont{Sivco}},
  \bibinfo{author}{\bibfnamefont{C.}~\bibnamefont{Sirtori}},
  \bibinfo{author}{\bibfnamefont{A.~L.} \bibnamefont{Hutchinson}},
  \bibnamefont{and} \bibinfo{author}{\bibfnamefont{A.~Y.} \bibnamefont{Cho}},
  \bibinfo{journal}{Science} \textbf{\bibinfo{volume}{264}},
  \bibinfo{pages}{553} (\bibinfo{year}{1994}).

\bibitem[{\citenamefont{Tonouchi}(2009)}]{qlaser3}
\bibinfo{author}{\bibfnamefont{M.}~\bibnamefont{Tonouchi}},
  \bibinfo{journal}{Terahertz Science and Technology}
  \textbf{\bibinfo{volume}{2}}, \bibinfo{pages}{90} (\bibinfo{year}{2009}).

\bibitem[{\citenamefont{Colson}(1985)}]{colson}
\bibinfo{author}{\bibfnamefont{W.~B.} \bibnamefont{Colson}},
  \bibinfo{journal}{Nucl.~Inst.~Meth.~Phys. A} \textbf{\bibinfo{volume}{237}},
  \bibinfo{pages}{1} (\bibinfo{year}{1985}).

\bibitem[{\citenamefont{Emma et~al.}(2004)\citenamefont{Emma, Bane, Cornacchia,
  Huang, Schlarb, Stupakov, and Walz}}]{Free2}
\bibinfo{author}{\bibfnamefont{P.}~\bibnamefont{Emma}},
  \bibinfo{author}{\bibfnamefont{K.}~\bibnamefont{Bane}},
  \bibinfo{author}{\bibfnamefont{M.}~\bibnamefont{Cornacchia}},
  \bibinfo{author}{\bibfnamefont{Z.}~\bibnamefont{Huang}},
  \bibinfo{author}{\bibfnamefont{H.}~\bibnamefont{Schlarb}},
  \bibinfo{author}{\bibfnamefont{G.}~\bibnamefont{Stupakov}}, \bibnamefont{and}
  \bibinfo{author}{\bibfnamefont{D.}~\bibnamefont{Walz}},
  \bibinfo{journal}{Phys. Rev. Lett} \textbf{\bibinfo{volume}{92}},
  \bibinfo{pages}{074801} (\bibinfo{year}{2004}).

\bibitem[{\citenamefont{Malkin and Fisch}(2007)}]{Fisch}
\bibinfo{author}{\bibfnamefont{V.~M.} \bibnamefont{Malkin}} \bibnamefont{and}
  \bibinfo{author}{\bibfnamefont{N.~J.} \bibnamefont{Fisch}},
  \bibinfo{journal}{Phys.~Rev.~Lett.} \textbf{\bibinfo{volume}{99}},
  \bibinfo{pages}{205001} (\bibinfo{year}{2007}).

\bibitem[{\citenamefont{Son et~al.}(2010)\citenamefont{Son, Ku, and
  Moon}}]{sonbackward}
\bibinfo{author}{\bibfnamefont{S.}~\bibnamefont{Son}},
  \bibinfo{author}{\bibfnamefont{S.}~\bibnamefont{Ku}}, \bibnamefont{and}
  \bibinfo{author}{\bibfnamefont{S.~J.} \bibnamefont{Moon}},
  \bibinfo{journal}{Phys.~Plasmas} \textbf{\bibinfo{volume}{17}},
  \bibinfo{pages}{114506} (\bibinfo{year}{2010}).

\bibitem[{\citenamefont{Son and Ku}(2009)}]{sonlandau}
\bibinfo{author}{\bibfnamefont{S.}~\bibnamefont{Son}} \bibnamefont{and}
  \bibinfo{author}{\bibfnamefont{S.}~\bibnamefont{Ku}},
  \bibinfo{journal}{Phys.~Plasmas} \textbf{\bibinfo{volume}{17}},
  \bibinfo{pages}{010703} (\bibinfo{year}{2009}).

\bibitem[{\citenamefont{Tabak et~al.}(1994)\citenamefont{Tabak, Hammer,
  Glinsky, Kruerand, Wilks, Woodworth, Campbell, Perry, and Mason}}]{tabak}
\bibinfo{author}{\bibfnamefont{M.}~\bibnamefont{Tabak}},
  \bibinfo{author}{\bibfnamefont{J.}~\bibnamefont{Hammer}},
  \bibinfo{author}{\bibfnamefont{M.~E.} \bibnamefont{Glinsky}},
  \bibinfo{author}{\bibfnamefont{W.~L.} \bibnamefont{Kruerand}},
  \bibinfo{author}{\bibfnamefont{S.~C.} \bibnamefont{Wilks}},
  \bibinfo{author}{\bibfnamefont{J.}~\bibnamefont{Woodworth}},
  \bibinfo{author}{\bibfnamefont{E.~M.} \bibnamefont{Campbell}},
  \bibinfo{author}{\bibfnamefont{M.~J.} \bibnamefont{Perry}}, \bibnamefont{and}
  \bibinfo{author}{\bibfnamefont{R.~J.} \bibnamefont{Mason}},
  \bibinfo{journal}{Physics of Plasmas} \textbf{\bibinfo{volume}{1}},
  \bibinfo{pages}{1626} (\bibinfo{year}{1994}).

\bibitem[{\citenamefont{Son and Fisch}(2005)}]{sonprl}
\bibinfo{author}{\bibfnamefont{S.}~\bibnamefont{Son}} \bibnamefont{and}
  \bibinfo{author}{\bibfnamefont{N.~J.} \bibnamefont{Fisch}},
  \bibinfo{journal}{Phys.~Rev.~Lett.} \textbf{\bibinfo{volume}{95}},
  \bibinfo{pages}{225002} (\bibinfo{year}{2005}).

\bibitem[{\citenamefont{Son and Fisch}(2004)}]{sonpla}
\bibinfo{author}{\bibfnamefont{S.}~\bibnamefont{Son}} \bibnamefont{and}
  \bibinfo{author}{\bibfnamefont{N.~J.} \bibnamefont{Fisch}},
  \bibinfo{journal}{Phys.~Lett.~A} \textbf{\bibinfo{volume}{329}},
  \bibinfo{pages}{16} (\bibinfo{year}{2004}).

\bibitem[{\citenamefont{Mangles et~al.}(2004)\citenamefont{Mangles, Murphy,
  Najmudin, Thomas, Collier, Dangor, Divall, Foster, Gallacher, Hooker
  et~al.}}]{monoelectron}
\bibinfo{author}{\bibfnamefont{S.~P.~D.} \bibnamefont{Mangles}},
  \bibinfo{author}{\bibfnamefont{C.~D.} \bibnamefont{Murphy}},
  \bibinfo{author}{\bibfnamefont{Z.}~\bibnamefont{Najmudin}},
  \bibinfo{author}{\bibfnamefont{A.~G.~R.} \bibnamefont{Thomas}},
  \bibinfo{author}{\bibfnamefont{J.~L.} \bibnamefont{Collier}},
  \bibinfo{author}{\bibfnamefont{A.~E.} \bibnamefont{Dangor}},
  \bibinfo{author}{\bibfnamefont{E.~J.} \bibnamefont{Divall}},
  \bibinfo{author}{\bibfnamefont{P.~S.} \bibnamefont{Foster}},
  \bibinfo{author}{\bibfnamefont{J.~G.} \bibnamefont{Gallacher}},
  \bibinfo{author}{\bibfnamefont{C.~J.} \bibnamefont{Hooker}},
  \bibnamefont{et~al.}, \bibinfo{journal}{Nature}
  \textbf{\bibinfo{volume}{431}}, \bibinfo{pages}{535} (\bibinfo{year}{2004}).

\bibitem[{\citenamefont{Tatarakis et~al.}(2003)\citenamefont{Tatarakis, Beg,
  Clark, Dangor, Edwards, Evans, Goldsack, Ledingham, Norreys, Sinclair
  et~al.}}]{ebeam}
\bibinfo{author}{\bibfnamefont{M.}~\bibnamefont{Tatarakis}},
  \bibinfo{author}{\bibfnamefont{F.~N.} \bibnamefont{Beg}},
  \bibinfo{author}{\bibfnamefont{E.~L.} \bibnamefont{Clark}},
  \bibinfo{author}{\bibfnamefont{A.~E.} \bibnamefont{Dangor}},
  \bibinfo{author}{\bibfnamefont{R.~D.} \bibnamefont{Edwards}},
  \bibinfo{author}{\bibfnamefont{R.~G.} \bibnamefont{Evans}},
  \bibinfo{author}{\bibfnamefont{T.~J.} \bibnamefont{Goldsack}},
  \bibinfo{author}{\bibfnamefont{K.~W.~D.} \bibnamefont{Ledingham}},
  \bibinfo{author}{\bibfnamefont{P.~A.} \bibnamefont{Norreys}},
  \bibinfo{author}{\bibfnamefont{M.~A.} \bibnamefont{Sinclair}},
  \bibnamefont{et~al.}, \bibinfo{journal}{Phys.~Rev.~Lett.}
  \textbf{\bibinfo{volume}{90}}, \bibinfo{pages}{175001}
  (\bibinfo{year}{2003}).

\bibitem[{\citenamefont{Carr et~al.}(2002)\citenamefont{Carr, Martin, McKinney,
  Jordan, Neil, and Williams}}]{Carr}
\bibinfo{author}{\bibfnamefont{G.~L.} \bibnamefont{Carr}},
  \bibinfo{author}{\bibfnamefont{M.~C.} \bibnamefont{Martin}},
  \bibinfo{author}{\bibfnamefont{W.~R.} \bibnamefont{McKinney}},
  \bibinfo{author}{\bibfnamefont{K.}~\bibnamefont{Jordan}},
  \bibinfo{author}{\bibfnamefont{G.~R.} \bibnamefont{Neil}}, \bibnamefont{and}
  \bibinfo{author}{\bibfnamefont{G.~P.} \bibnamefont{Williams}},
  \bibinfo{journal}{Nature} \textbf{\bibinfo{volume}{420}},
  \bibinfo{pages}{153} (\bibinfo{year}{2002}).

\bibitem[{\citenamefont{Esarey et~al.}(2005)\citenamefont{Esarey, van. Tilborg,
  Michel, Schroeder, Toth, Geddes, and Shadwick}}]{Thz}
\bibinfo{author}{\bibfnamefont{E.}~\bibnamefont{Esarey}},
  \bibinfo{author}{\bibfnamefont{J.}~\bibnamefont{van. Tilborg}},
  \bibinfo{author}{\bibfnamefont{P.~A.} \bibnamefont{Michel}},
  \bibinfo{author}{\bibfnamefont{C.~B.} \bibnamefont{Schroeder}},
  \bibinfo{author}{\bibfnamefont{C.}~\bibnamefont{Toth}},
  \bibinfo{author}{\bibfnamefont{C.~G.~R.} \bibnamefont{Geddes}},
  \bibnamefont{and} \bibinfo{author}{\bibfnamefont{B.~A.}
  \bibnamefont{Shadwick}}, \bibinfo{journal}{IEEE Transactions on Plasma
  Science} \textbf{\bibinfo{volume}{33}}, \bibinfo{pages}{8}
  (\bibinfo{year}{2005}).

\bibitem[{\citenamefont{Leemans et~al.}(2004)\citenamefont{Leemans, van.
  Tilborg, Faure, R.Geddes, Toth, Schroeder, Esarey, Fubiani, and
  Dugan}}]{Thz2}
\bibinfo{author}{\bibfnamefont{W.~P.} \bibnamefont{Leemans}},
  \bibinfo{author}{\bibfnamefont{J.}~\bibnamefont{van. Tilborg}},
  \bibinfo{author}{\bibfnamefont{J.}~\bibnamefont{Faure}},
  \bibinfo{author}{\bibfnamefont{C.~G.} \bibnamefont{R.Geddes}},
  \bibinfo{author}{\bibfnamefont{C.}~\bibnamefont{Toth}},
  \bibinfo{author}{\bibfnamefont{C.~B.} \bibnamefont{Schroeder}},
  \bibinfo{author}{\bibfnamefont{E.}~\bibnamefont{Esarey}},
  \bibinfo{author}{\bibfnamefont{G.}~\bibnamefont{Fubiani}}, \bibnamefont{and}
  \bibinfo{author}{\bibfnamefont{G.}~\bibnamefont{Dugan}},
  \bibinfo{journal}{Physics of Plasmas} \textbf{\bibinfo{volume}{11}},
  \bibinfo{pages}{2899} (\bibinfo{year}{2004}).

\bibitem[{\citenamefont{Stix}(1962)}]{stix}
\bibinfo{author}{\bibfnamefont{T.~H.} \bibnamefont{Stix}},
  \emph{\bibinfo{title}{The theory of plasma waves}}
  (\bibinfo{publisher}{McGraw-Hill}, \bibinfo{year}{1962}).

\bibitem[{\citenamefont{Son and Moon}(2012)}]{songamma}
\bibinfo{author}{\bibfnamefont{S.}~\bibnamefont{Son}} \bibnamefont{and}
  \bibinfo{author}{\bibfnamefont{S.~J.} \bibnamefont{Moon}},
  \bibinfo{journal}{Phys.~Plasmas} \textbf{\bibinfo{volume}{19}},
  \bibinfo{pages}{063102} (\bibinfo{year}{2012}).

\bibitem[{\citenamefont{Geddes et~al.}(2004)\citenamefont{Geddes, Toth, van.
  Tilborg, Esarey, Schroeder, Bruhwiler, Nieter, Cary, and Leemans}}]{wake}
\bibinfo{author}{\bibfnamefont{C.~G.~R.} \bibnamefont{Geddes}},
  \bibinfo{author}{\bibfnamefont{C.}~\bibnamefont{Toth}},
  \bibinfo{author}{\bibfnamefont{J.}~\bibnamefont{van. Tilborg}},
  \bibinfo{author}{\bibfnamefont{E.}~\bibnamefont{Esarey}},
  \bibinfo{author}{\bibfnamefont{C.~B.} \bibnamefont{Schroeder}},
  \bibinfo{author}{\bibfnamefont{D.}~\bibnamefont{Bruhwiler}},
  \bibinfo{author}{\bibfnamefont{C.}~\bibnamefont{Nieter}},
  \bibinfo{author}{\bibfnamefont{J.}~\bibnamefont{Cary}}, \bibnamefont{and}
  \bibinfo{author}{\bibfnamefont{W.~P.} \bibnamefont{Leemans}},
  \bibinfo{journal}{Nature} \textbf{\bibinfo{volume}{431}},
  \bibinfo{pages}{538} (\bibinfo{year}{2004}).

\bibitem[{\citenamefont{Malka et~al.}(2008)\citenamefont{Malka, Faure, Gauduel,
  Lefebvre, Rousee, and Phuoc}}]{wake2}
\bibinfo{author}{\bibfnamefont{V.}~\bibnamefont{Malka}},
  \bibinfo{author}{\bibfnamefont{J.}~\bibnamefont{Faure}},
  \bibinfo{author}{\bibfnamefont{Y.~A.} \bibnamefont{Gauduel}},
  \bibinfo{author}{\bibfnamefont{E.}~\bibnamefont{Lefebvre}},
  \bibinfo{author}{\bibfnamefont{A.}~\bibnamefont{Rousee}}, \bibnamefont{and}
  \bibinfo{author}{\bibfnamefont{K.~T.} \bibnamefont{Phuoc}},
  \bibinfo{journal}{Nature Physics} \textbf{\bibinfo{volume}{4}},
  \bibinfo{pages}{447} (\bibinfo{year}{2008}).

\end{thebibliography}

\end{document}